\documentclass[aps,prl,showpacs,amsmath,amssymb,lengthcheck]{revtex4}

\usepackage{epsfig}

\begin{document}
\title{Normal-Superfluid Interface Scattering For Polarized
  Fermion Gases}

\author{{\small Bert Van Schaeybroeck and Achilleas Lazarides}}
\address{{\small\it Laboratorium voor Vaste-Stoffysica en
Magnetisme,}\\{\small\it Celestijnenlaan 200 D, Katholieke
Universiteit Leuven}, {\small\it B-3001 Leuven, Belgium.}}

\date{\small\it \today}
\newcommand{\xip}{\ensuremath{\xi_p}}

\begin{abstract}
  We argue that, for the recent experiments with imbalanced fermion gases, a
  temperature difference may occur between the normal (N) and the gapped
  superfluid (SF) phase. Using the mean-field formalism, we study particle
  scattering off the N-SF interface from the deep BCS to the unitary regime.
  We show that the thermal conductivity across the interface drops
  exponentially fast with increasing $h/k_B T$, where $h$ is the chemical
  potential imbalance. This implies a blocking of thermal equilibration
  between the N and the SF phase. We also provide a possible mechanism for the
  creation of gap oscillations (FFLO-like states) as seen in recent
  studies on these systems.
\end{abstract}


\maketitle \textit{Introduction} --- An electron approaching a
normal-superconducting (N-SC) interface from the normal side with energy
$E<\Delta$, where $\Delta$ is the superconducting gap, has insufficient energy
to excite quasiparticles inside the SC and is therefore reflected. In a SC,
the relation $\Delta\ll E_F$ (where $E_F$ is the Fermi energy) constrains the
momentum transfer during interfacial scattering to be much lower than $2\hbar
k_F$, required for normal (specular) reflection. Consequently, Andreev
reflection occurs: the electron pairs with another electron of opposite
momentum, forming a Cooper pair. A ``reflected" hole is left on the N side,
which follows the time-reversed path of the incident electron. Andreev used
this process to describe the unusual heat conductivity found in N-SC junctions
and in the intermediate state of superconductors~\cite{andreev}. To date, the
transport properties of such structures remain the subject of intense
research~\cite{beenakker,blonder}. On the other hand, recent
experiments~\cite{partridge,zwierlein} have probed superfluidity in ultracold
fermionic mixtures, where the possibility arises of having different chemical
potentials for each species and controllable interspecies interactions.  An
accurate theoretical prediction of the observed density profiles is still not
available, despite intense theoretical
activity~\cite{bedaque,chevy,sheehy,parish,desilva,gubbels,castorina}. Whereas the incorporation of a normal-superfluid (N-SF) interface tension may
settle this question~\cite{desilva,gubbels} for the Rice experiments, the
interpretation of the MIT experiments remains unclear. A recently proposed 
theory~\cite{gubbels} appears to explain all observed features, although the
temperatures required for agreement with data from the MIT experiments are
five times higher than observed.

In the following, we argue that the presence of a N-SF interface is likely
to block thermal equilibration for these experiments, possibly inducing a
temperature difference between the two phases. Incorporating such a temperature
difference in the existing models may provide the key to a complete understanding of
the experiments.

In this Letter we focus on two question, to wit: 1) What are
the possible reflection and transmission mechanisms at a N-SF
interface? The main features distinguishing this system from a SC
are the difference in the chemical potentials and the variation of
the gap from $\Delta\ll E_F$ to $\Delta\sim E_F$ as one tunes the
interactions from deep BCS to the unitary regime. 2) What is the
relevance of these findings for recent experimental and
theoretical works? 

We find a rich variety of interfacial scattering processes, depending on the
energy and perpendicular momentum of the incoming particle. The scattering of
the particles off the interface results in a striking decrease in the thermal
conductivity for temperatures $ T\lesssim 0.05\, T_F$ (at unitarity). We also argue that upon
Andreev reflection, particles and holes interfere so as to cause gap
oscillations near the interface, which are observed in recent numerical
studies~\cite{castorina}.  We postpone the description of calculational
details and the case of unequal fermion masses~\cite{lazarides}.

The system under study consists of two fermionic species $a$ and $b$ with
equal masses $m$ and chemical potentials $\mu_{i}^0$ for $i=a,\,b$, trapped by
a potential $\mathrm{V}(\mathbf{r})$; this gives rise to an effective chemical
potential $\mu_{i}(\mathbf{r})=\mu_{i}^0-\mathrm{V}(\mathbf{r})$ (henceforth
denoted by $\mu_{i}$). The Bogoliubov-de Gennes (BdG) or
Blonder-Tinkham-Klapwijk equations~\cite{degennes, blonder} give a
satisfactory description of this system. Denoting the particle-like and
hole-like eigenfunctions of species $i=a,\,b$ by $u_{i,\mathbf{k}}$ and
$v_{i,\mathbf{k}}$ respectively and following the standard procedure, one
finds:
\begin{align}\label{Bdg}
  \left[ \begin{array}{cc}
\mathcal{H}_a & \Delta\\
\Delta^* & -\mathcal{H}_b
\end{array} \right]
 \left[ \begin{array}{c}
u_{a,\textbf{k}}\\
v_{b,\textbf{k}}
\end{array} \right]
= E_{\textbf{k}} \left[ \begin{array}{c}
u_{a,\textbf{k}}\\
v_{b,\textbf{k}}
\end{array} \right],
\end{align}
with $\mathcal{H}_{a,b}=-\boldsymbol{\nabla}^2/2m-\mu\mp h$ where
$\mu=(\mu_a+\mu_b)/2>0$, $h=(\mu_a-\mu_b)/2>0$ and we set
$\hbar=k_B=1$. One obtains the second set of BdG Eqs.~by
interchanging $a$ and $b$. As seen from Eqn.~\eqref{Bdg}, incoming
$a$-particles are coupled to $b$-holes and vice versa. Now, the
prerequisite for an interface to be present between a SF with
symmetrical $a$ and $b$ densities and an asymmetrical N phase is
the existence of a first-order transition between the two phases.
In the experiments, if a SF is found at the trap center, $\mu$
decreases upon approaching the trap boundary and probes the
$(\mu_a,\mu_b)$ phase diagram, possibly inducing a crossing of the
N-SF phase boundary. This happens when locally there occurs a
balance between the energy gained by creating, on the one hand, a
gap and on the other hand, a density difference~\footnote{Strictly
speaking, this is only
  valid if the trapping length is much longer than the
  interfacial length.}. At $T=0$, this local coexistence
condition is well-approximated by the Clogston
limit~\cite{lazarides}:
\begin{align}\label{coexistence}
  \Delta=\sqrt{2}\,h.
\end{align}
For fixed interaction parameter $k_F a$, where $a$ is the scattering length,
the first-order transition persists at finite temperature up to a tricritical
point~\cite{parish,gubbels}. The Rice experiments are indeed accurately
described using a theory incorporating a first-order
transition~\cite{desilva,gubbels,partridge, chevy}; the MIT results are also
suggestive of a first-order transition.

To model the N-SF interface, it is natural to start with a
geometry wherein the $x=0$ plane separates the N from the SF
region:
\begin{align}\label{gap}
  \Delta(\mathbf{r})=\Theta(x) \Delta,
\end{align}
with $\Theta$ the Heaviside function.
\begin{figure}
   \epsfig{figure=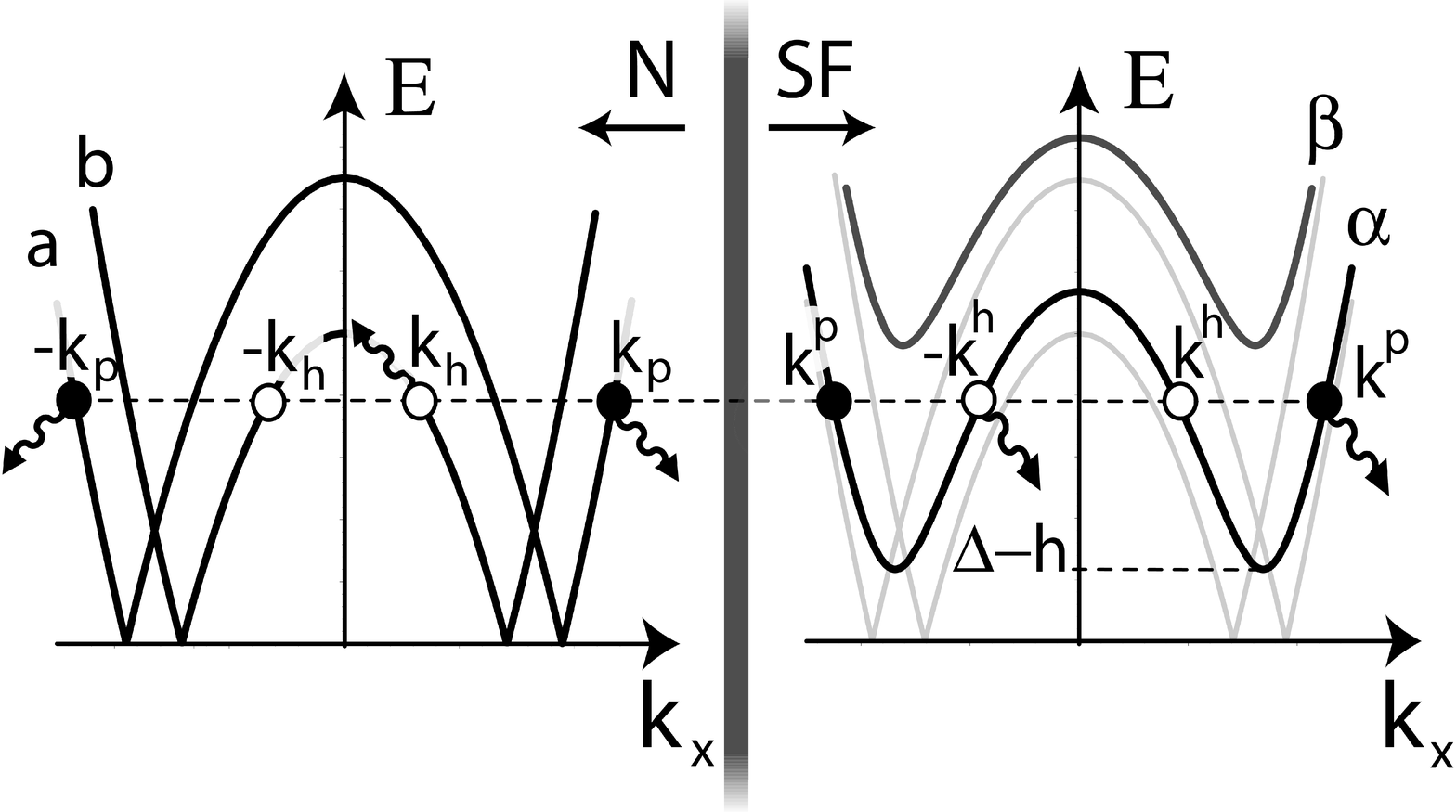,angle=0, height=140pt}
       \caption{
       The N-SF interface (thick vertical line) with the $a,\,b$ spectra on the N
       side and the gapped
       $\alpha,\,\beta$ spectra in the SF. The long dashed line
       cuts the spectra at particle-like (filled dots) and
       hole-like (empty dots) quasiparticle states, all having the same
       energy. An incoming $a$-particle (curly arrow) with momentum $k_p$ and
       energy $E>\Delta-h$ has up to four scattering channels: the Andreev
       reflected $k_h$ hole, the specularly reflected $-k_p$ particle and
       the transmitted hole-like $-k^h$ and particle-like $k^p$ states.
       \label{fig1}
     }
\end{figure}
This choice allows one to find the general solutions of the BdG
Eqs.~\eqref{Bdg} in the N and SF phase. They read:
{\small
  \begin{align}\label{general}
    \left( \begin{array}{c}
        u_a^n \\
        v_b^n\\
      \end{array} \right)
    &=\sum_{\mathbf{k},\pm} e^{i\mathbf{k}_\|\cdot \mathbf{r}}
    \left( \begin{array}{c}
        U_{\mathbf{k},\pm}^{p,n}\,\, e^{\pm i k_px}\\
        V_{\mathbf{k},\pm}^{h,n}\,\, e^{\pm i k_hx}
      \end{array} \right),\\
    \left( \begin{array}{c}
        u_a^s \\
        v_b^s\\
      \end{array} \right)
    &=\sum_{\mathbf{k},\pm}e^{i\mathbf{k}_\|\cdot \mathbf{r}} \left[
      \left(
        \begin{array}{c}
          U_{\mathbf{k},\pm}^{h,s}\\
          V_{\mathbf{k},\pm}^{h,s}
        \end{array} \right)
      e^{\pm i k^hx} + \left( \begin{array}{c}
          U_{\mathbf{k},\pm}^{p,s}\\
          V_{\mathbf{k},\pm}^{p,s}
        \end{array} \right)
      e^{\pm i k^px}\right].\nonumber
  \end{align}
}
Here, the sub- and superscripts $n,\, s,\, p$ and $h$ denote
normal, superfluid, particle-like and hole-like respectively. We
split the vectors $\mathbf{k}$ into their components parallel to
the wall $\mathbf{k}_\|=(0,k_y,k_z)$ and the $x$-component $k_p$
which relates to the $k^{p,h}_h$ by the BdG Eqs.:
\begin{subequations}
  \begin{align}
    k_h=&\sqrt{k_p^2 -4m (E+h)},\label{khone}\\
    k^{p,h} =&\sqrt{k_p^2 -2m\xi_\mp}\label{khtwo},
  \end{align}
\end{subequations}
\begin{figure}
   \epsfig{figure=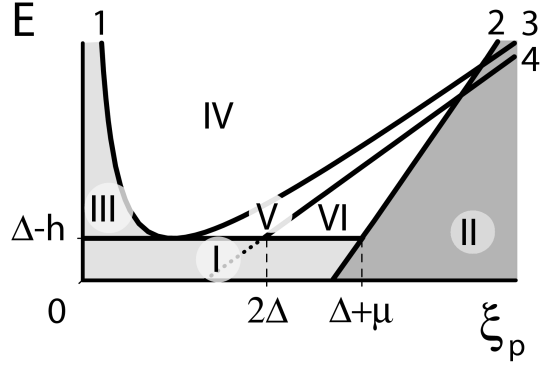,angle=0, height=140pt}
   \caption{The various scattering regions for an $a$-particle incident on
     the interface from the N side, as a function of its energy $E$ and
     $\xip=k^2_p/2m$ ($k_p$ is its momentum along the $x$-axis). The
     heavily-shaded region is energetically forbidden, while complete
     reflection occurs in the lightly-shaded regions. Particles in region~VI
     may scatter to all states indicated in Fig.~\ref{fig1} by curly arrows.
     Above line 4 (regions~IV and V), Andreev reflections do not occur, and
     above curve 3 (region~IV), hole-like excitations are also impossible. The
     numbered curves are 1:~$\xip=\xi_-$, 2:~$\xip=\mu_a+E$, 3:~$\xip=\xi_+$
     and 4:~$\xip=2(E+h)$.\label{fig2}}
\end{figure}
where $\xi_\pm=E+h\pm\sqrt{(E+h)^2-\Delta^2}$. The amplitudes $U$
and $V$ from solutions~\eqref{general} are determined by matching
the wavefunctions and their derivatives at $x=0$~\cite{demers}.
Figure~\ref{fig1} shows the quasiparticle spectra in both the N
region (left panel) and the SF region (right panel). In the former
one recognizes two approximately linear energy branches $a$ and
$b$ near the Fermi surfaces, while in the latter the $\alpha$ and
$\beta$ spectra are gapped by $\Delta-h$ and $\Delta+h$. Since we
take $\mu_a>\mu_b$, the states belonging to the $\alpha$ (lower)
spectrum are the ones composed of $a$-particles and $b$-holes.

\textit{Scattering regimes} --- Consider an incoming
$a$-particle from the N side with energy $E$ and momentum
$\mathbf{k}$; it is indicated in Fig.~\ref{fig1} by a curly arrow
and its motion is completely characterized by $E$ and
$\xip=k_p^2/2m$. The crossover of either of the $k_{h}^{p,h}$ or
$\mathbf{k}_\|$ from real to imaginary (or vice versa) signifies a
change in the scattering mechanism. The regions in the $E-\xip$
plane corresponding to the various such scattering processes are
shown in Fig.~\ref{fig2}; we now briefly describe them. If
$E<\Delta-h$, the incoming particle has insufficient energy to
excite quasiparticles inside the SF; it is completely reflected as
a superposition of a particle and a hole. This region is
labelled~I in Fig.~\ref{fig2}. The situation where $E<\xip-\mu_a$
is physically forbidden for obvious reasons; it is labelled~II.
Next, consider energies above the threshold $\Delta-h$. For
$\xip<\xi_-$ (region~III), the incoming particle can excite
neither particle-like nor hole-like quasiparticles in the SF and
must therefore undergo complete reflection, even though it appears
to have sufficient energy for transmission. The situation is akin
to that of a quantal particle incident on a potential step; if the
angle of incidence $\theta$ exceeds a critical angle, the particle
is reflected. Here, the critical angle is $\theta_c^-$ with
$\tan\theta_c^\pm=\sqrt{(E+h+\mu)/\xi_\pm-1}$, in which
$\theta_c^-<\theta$ corresponds to $\xip<\xi_-$. For angles
$\theta$ satisfying $\theta_c^+<\theta<\theta_c^-$ (region~IV), or
$\xi_-<\xip<\xi_+$, particle-like states may be excited in the SF,
in contrast to hole-like states.  When $\xip<2(E+h)$ (regions~IV
and V), there are no reflected holes; that is, Andreev reflection
does not occur. Thus, in region~V, both particle-like and
hole-like excitations are present in the SF, but Andreev
reflection is impossible.  Finally, in region~VI, both
particle-like and hole-like excitations, as well as Andreev and
normal reflection are allowed. To summarize, the lightly-shaded
regions in Fig.~\ref{fig2} describe $a$-particles which undergo
complete reflection, while the rightmost, heavily-shaded region is
unphysical. Only incoming particles in the unshaded regions may
excite quasiparticles inside the SF. As for incoming holes, they
are only transmitted in a region analogous to~II in the diagram.
Arguments very similar to the preceding, applied to incoming
$b$-particles and $a$-holes (therefore the $\beta$-channel), lead
to a diagram identical to that of Fig.~\ref{fig2} but with
$h\rightarrow -h$.

The $\xi_p$-$E$ diagrams of the deep BCS and the unitary regime
have a different topology. In the deep BCS regime, the relation
$2\Delta\ll \Delta+\mu$ implies that region~VI of Fig.~\ref{fig2}
is by far the most important. Quasiparticle reflections occur
mostly via the Andreev mechanism which involves particle, but not
energy, transport across the interface. In contrast, at unitarity
where $2\Delta> \Delta+\mu$ (since $\Delta\approx 1.16\mu$), regime
VI does not even exist. This means that quasiparticles with energy
above the threshold $\Delta-h$ cannot undergo Andreev reflection
but only normal (specular) reflection, in which neither particles
nor energy are carried across the interface.

\textit{Thermal conductivity $\kappa$} --- In order to study the
transport properties, we seek to relate the amplitudes $U$ and $V$
of Eqs.~\eqref{general} to the transport coefficients, based on a
conservation law for the current. From the BdG Eqs.~it readily
follows that the density $\rho_\alpha(\mathbf{r})=|u_a|^2+|v_b|^2$
and the quasiparticle current
\begin{align}
  \mathbf{j}_\alpha&=-\frac{i}{2m}\left[u_a^*\boldsymbol{\nabla}
    u_a-u_a\boldsymbol{\nabla} u_a^*-v_b^*\boldsymbol{\nabla}
    v_b+v_b\boldsymbol{\nabla} v_b^*\right]\nonumber
\end{align}
satisfy the continuity equation. The transmission coefficient
$S(E,\xi_p)$ of an incoming particle of energy $E$ and momentum
$\mathbf{k}$ is defined as the ratio of the transmitted to the
incoming current along the $x$-axis. $S$ vanishes in regions~I, II
and III whereas for energies slightly above the transmission
threshold $E\approx \Delta-h$ (for regions~V and VI),
\begin{align}\label{squarroot}
S(E,\xi_p)\propto \sqrt{E-(\Delta-h)},
\end{align}
similarly to the case of particles scattering from a Hartree
potential of height $\Delta-h$. Next, one can write the heat flux
through the interface and to the $i=\alpha,\,\beta$ branch as:
\begin{align}\label{flux}
  W_{x,i}&=\frac{m}{4\pi^2}\sum_{s=p,h}\int
  \text{d}\xi_{s}\int\text{d}E\, E
  f(E)S(E,\xi_s),
\end{align}
where $f(E)$ is the Fermi-Dirac distribution and the integration
is performed over the $\xip$-$E$ and $\xi_h$-$E$ planes (with
$\xi_h=k_h^2/2m$). In equilibrium, the N-SF flux is balanced by an
equal SF-N flux. A small temperature bias on the N side will
induce a net heat flow $Q$. By the Kapitza approach, the latter is
$Q=\kappa\,\delta T$ with the heat conductivity $\kappa=\partial
(W_{x,\alpha}+W_{x,\beta})/\partial T$.

\begin{figure}
   \epsfig{figure=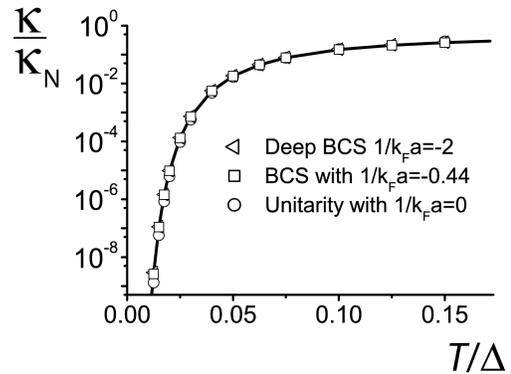,angle=0, height=150pt}
       \caption{The thermal conductivity across the N-SF interface $\kappa$
         divided by the normal-phase
       conductivity $\kappa_N$ against $T/\Delta$ for the unitary, BCS and deep
       BCS cases. For $T\lesssim 0.1\Delta$, $\kappa/\kappa_N$ drops
       dramatically (notice the logarithmic scale). The curve represents the analytical result
       obtained by the use of the Andreev approximation, Eqn.~\eqref{andreev}.
       At unitarity, $\Delta=0.69T_F$, in the BCS case $\Delta=0.25 T_F$
       and in the deep BCS case $\Delta=0.002\, T_F$.
       \label{fig3}
     }
\end{figure}

We have analytically calculated the transmission coefficients for
all regions of Fig.~\ref{fig2}. To fix the values of $h,\,\mu$ and
$T_F=k_F^2/2m=(3\pi^2n)^{2/3}/2m$ for given $k_F a$ ($n$ is the
density in the SF), we have used the gap and number equations, as
well as the exact $T=0$ coexistence condition~\cite{lazarides}.
The resulting ratio of $\kappa$ to the conductivity in the N
phase, $\kappa_N$, is shown in Fig.~\ref{fig3} as a function of
$T/\Delta$. We find that $\kappa/\kappa_N$ decreases drastically
below $T\approx 0.1 \Delta$~\footnote{The normal-state
conductivity $\kappa_N$ vanishes linearly
  with $T$.}.
Remarkably, $\kappa/\kappa_N$ is almost independent of the
interaction parameter $k_F a$. This can be understood as follows.
For low enough temperature, $T\ll\Delta$, only incident particles
(and holes) with energies slightly above the threshold $E\approx
\Delta-h$ contribute to the heat conductivity because of their low
statistical weight $f\propto e^{ -(\Delta-h)/T}$ appearing in
eqn.~\eqref{flux}. Furthermore, relation~\eqref{squarroot}
shows that $S$ displays the same behavior in regions
V~and~VI. The main effect of varying $1/k_F a$ from large and
negative (deep BCS) to zero (unitarity) is to move the boundary
between regions V~and~VI to the right. Both the strong exponential
decay with $\Delta/T$ and the square root dependence of $S$ on
$E-(\Delta-h)$ are unaffected by the variation of $1/k_F a$, hence
the very similar behavior of $\kappa/\kappa_N$ for all regimes
under study.

These considerations are exemplified by the following calculation
in the deep BCS regime, where the Andreev approximation may be
used to obtain an analytical expression for $\kappa$. The
approximation is to equate all the $k^{p,h}_{p,h}$ wave vectors
and take the temperature to satisfy $T\ll \Delta\ll \mu$. We
obtain
\begin{align}\label{andreev}
\kappa_{BCS}=2\sum_{\pm}\frac{\sqrt{2\pi}m\mu}{4\pi^2}
\frac{e^{-(\Delta\pm h)/T}(\Delta\pm h)^2}{\sqrt{\Delta T}},
\end{align}
which amounts to Andreev's result when $h=0$~\cite{andreev}.  The
summation denotes the excitations of the $\alpha$ (-) and $\beta$
(+) states in the SF. The energy carried by the $\beta$ branch is
a factor $e^{-2h/T}$ lower than that of the $\alpha$ branch and
due to coexistence $h\sim \Delta\gg T$, it can be neglected.
Equation~\eqref{andreev} demonstrates the exponential decay of
$\kappa$ with decreasing temperature. Although the Andreev
approximation is invalid for the aforementioned
experiments~\cite{partridge,zwierlein}, formula~\eqref{andreev} as
plotted in Fig.~\ref{fig3}, provides a good estimate of $\kappa$,
even beyond the BCS regime~\footnote{In the plot we include the
next order term (in $T/\Delta$) in Eqn.~\eqref{andreev}.}.

\textit{FFLO states} --- Andreev reflection may also produce
effects similar to Fulde-Ferrell-Larkin-Ovchinnikov states
(FFLO) as seen in Ref.~\onlinecite{castorina}. For the case of a
SC, McMillan calculated~\cite{mcmillan} the first corrections to
the gap function~\eqref{gap} due to self-consistency and found, at
$T=0$, a decaying oscillation of $\Delta(x)$ on the N side. This
stems from the phase difference $\pm(k_p-k_h)$ between the
wavefunctions of incoming particles (holes) with momentum $k_p\,
(k_h)$ and energy $E\lesssim\Delta$, and their Andreev reflected
holes (particles) of momentum $k_h \,(k_p)$; the wavefunctions of
these determine the gap profile by
$\Delta(x)\propto\sum_\mathbf{k} u_{a,\mathbf{k}}(x)
v_{b,\mathbf{k}}^*(x)$.  We speculate that this also happens
beyond the BCS regime. In particular, the numerically observed $T=0$
FFLO states for trapped gases at unitarity may be a consequence of
the presence of the N-SF interface~\cite{castorina}. Note that the
FFLO state is not thermodynamically stable for a homogeneous system at
unitarity~\cite{sheehy}.

\textit{Discussion} --- In a SC, the heat conductivity also has a
lattice component, which dominates the electronic component at low
temperatures; the absence of such a component in the system under
study makes the decrease of the conductivity more significant. 
Addition of the Hartree-Fock potentials amounts to a change of the
coexistence condition~\eqref{coexistence} and a mere horizontal
shift of lines 1 through 4 in Fig.~\ref{fig2}, thereby preserving the
energy gap (region~I) and thus the main conclusions of our paper.
We expect that improvements on our approximations, which are the
use of the one-channel model, the gap profile~\eqref{gap}, and the
temperature dependence which is solely contained in the
Fermi-Dirac function, will affect the results only quantitatively
and mostly at unitarity. Finally, note that the values we used for
$h,\,\mu$ and $E_F$ are those at the interface. Estimating that
$T_F$ at unitarity (see caption of Fig.~\ref{fig3}) is equal the reported
$T_F$~\cite{zwierlein, partridge}, one concludes that the drop of
conductivity sets in at $T=0.05\, T_F$, from which follows that
our arguments indeed apply to the current experiments.

\textit{Conclusion} --- Scattering of quasiparticles off the N-SF interface,
as summarized in Fig.~\ref{fig2}, gives rise to the following effects.  The
thermal conductivity across the interface decreases rapidly with decreasing
temperature at the experimentally realized temperatures, and is described by
equation~\eqref{andreev}. This implies that thermalization slows down, and a
temperature difference across the interface can appear; the incorporation of
this could lead recent models to better agreement with current experiments. We
call for such incorporation, as well as for a separate temperature measurement
of the N and SF phase since, as we have argued, they may not be the same. In
addition, we argue that reflections of quasiparticles off the interface can
induce modifications to (BCS regime), or even cause (at unitarity) gap
oscillations, usually identified as FFLO state.

\textit{Acknowledgement} --- We acknowledge partial support by
project FWO G.0115.06; AL and BVS is supported by GOA/2004/02; BVS is
supported by IUAP P5/01. It is a pleasure to thank Henk Stoof and
Koos Gubbels for discussions and useful suggestions and Joseph
Indekeu for a careful reading of the manuscript.

\end{document}